\documentclass[twocolumn,showpacs,floatfix,superscriptaddress, showkeys,prc]{revtex4-2}%

\usepackage{mathrsfs}
\usepackage{graphicx}
\usepackage{dcolumn}
\usepackage{bm}
\usepackage[colorlinks,linkcolor=blue,citecolor=blue]{hyperref}
\usepackage{amsmath, amssymb}
\usepackage{CJK}
\usepackage{multirow}
\usepackage[section]{placeins}
\usepackage{tikz}

\newcommand{\svec}[1]{\boldsymbol{#1}}

\newcommand{\cals}[1]{{\mathcal #1}}
\newcommand{\ff}[1]{\frac{1}{#1}}

\newcommand{\lrlc}[1]{\left|#1\right>}
\newcommand{\lrcl}[1]{\left<#1\right|}

\usepackage{ulem}

\newcommand{\delete}{\bgroup\markoverwith{\textcolor{red}{\rule[0.5ex]{2pt}{1pt}}}\ULon}

\begin{document}
\title{Time-dependent Relativistic Hartree-Fock model with spherical symmetry}

\author{Jing Geng}
\affiliation{Frontier Science Center for Rare isotope, Lanzhou University, Lanzhou 730000, China}
\author{Zhi Heng Wang}
\affiliation{Frontier Science Center for Rare isotope, Lanzhou University, Lanzhou 730000, China}
\author{Peng Wei Zhao}
\affiliation{State Key Laboratory of Nuclear Physics and Technology, School of Physics, Peking University, Beijing 100871, China}
\author{Yi Fei Niu}
\affiliation{Frontier Science Center for Rare isotope, Lanzhou University, Lanzhou 730000, China}
\author{Haozhao Liang}
\affiliation{Department of Physics, Graduate School of Science, The University of Tokyo, Tokyo 113-0033, Japan}
\author{Wen Hui Long}\email{longwh@lzu.edu.cn} 
\affiliation{Frontier Science Center for Rare isotope, Lanzhou University, Lanzhou 730000, China}
\affiliation{Joint Department for Nuclear Physics, Lanzhou University and Institute of Modern Physics, CAS, Lanzhou 730000, China}

\begin{abstract}

This work establishes the time-dependent relativistic Hartree-Fock (TD-RHF) model with spherical symmetry for the first time. The time-dependent integro-differential Dirac equations are solved by expanding Dirac spinors on the spherical Dirac Woods-Saxon (DWS) basis. The numerical verification demonstrates the high conservation qualities for both the total binding energy and the particle number, as well as the time-reversal invariance of the system, which ensures the precision and reliability of the newly developed TD-RHF model. Subsequently, the isoscalar giant monopole resonance (ISGMR) mode of $^{208}$Pb is investigated using the RHF Lagrangian PKO1. The constrained energy of the ISGMR calculated by PKO1 is found to be in close agreement with the experimental data, and the strength function is similar to the results given by the relativistic Hartree-Fock plus random phase approximation. Based on the advantage of the TD-RHF model in avoiding complicated calculations of the residual interactions, the ISGMR mode of $^{208}$Pb is calculated by twelve relativistic effective Lagrangians. The results indicate that the value of the incompressibility of nuclear matter $K_\infty$ constrained by relativistic effective Lagrangians is in the range of $237\sim246$ MeV, which is lower than the previous investigations based on the relativistic models.

\end{abstract}


\maketitle


\section{Introduction}

Giant resonances (GRs) are the most conspicuous manifestation of nuclear collective motion, encompassing highly collective nuclear excitations that involve a substantial fraction of nucleons. The GRs provide access to extract crucial information about the nuclear equation of state (EoS), a fundamental ingredient for the description of supernova explosions, neutron stars and heavy ion collisions \cite{Roca-Maza2018PPNP101.96, Burgio2021PPNP120.103879, Oertel2017RMP89.015007, Perego2022PRL129.032701}. The incompressibility of nuclear matter, $K_\infty$, is a crucial parameter of the EoS \cite{Garg2018PPNP101.55}. The value of $K_\infty$ near the saturation point of nuclear matter can be confined by experimental measurements of the energies of the isoscalar giant monopole resonances (ISGMR) of nuclei, known as the breathing mode. The ISGMR characterises a mode that the nucleus alternately shrinks and expands, with the radial density undergoing an oscillation around the equilibrium point. Given such a picture, the linear correlation between the ISGMR energies and $K_\infty$, which can be verified by systematical model calculations, serves as a powerful tool to confine $K_\infty$ \cite{Blaizot1995NPA591.435}.

In verifying the linear correlation between ISGMR and the incompressibility $K_\infty$, one of the most powerful theoretical recipes is the random phase approximation (RPA) and its subsequent extension, the quasiparticle random phase approximation (QRPA) \cite{Colo2022arXiv2201.04578}. According to self-consistent (Q)RPA calculations based on the non-relativistic density functional theory, the value of $K_\infty$ was confined in a range of $220\sim235$ MeV \cite{Colo2004PAN67.1759}. However, the studies with the relativistic mean field (RMF) models indicated larger $K_\infty$ values ranging from 250 to 270 MeV \cite{Ma2001NPA686.173, Vretenar2003PRC68.024310}. It should be noted that such large value given by the RMF+RPA model was primarily based on earlier effective Lagrangians, such as NL$i$ ($i$=1,2,3) \cite{Reinhard1986ZPA323.13, Lee1986PRL57.2916, Lalazissis1997PRC55.540}, NLSH \cite{Sharma1993PLB312.377}, TM1 \cite{Sugahara1994NPA579.557} and HS \cite{Horowitz1981NPA368.503}. With the advent of the newer effective Lagrangians DD-ME2 \cite{Lalazissis2005PRC71.024312} and DD-PC1 \cite{Niksic2008PRC78.034318}, it was indicated that the value of $K_\infty$ predicted by the relativistic models can be lower in comparison to the earlier investigations \cite{Ma2001NPA686.173, Vretenar2003PRC68.024310}. Therefore, under the relativistic scheme, it is deserved to test the constraint on $K_\infty$ more systematically, by considering a larger range of popular relativistic effective Lagrangians. 

In the (Q)RPA framework, the residual interactions, which are obtained from the second variation of the nuclear energy functional with respect to the one-body nucleonic density, shall be handled with great care. On the market, the relativistic effective Lagrangians can be divided into two main categories: the meson-exchange and point-coupling versions, regarding of the range of the effective nuclear force. More than that, such categories may be even subdivided with respect to the modelling of the effective nuclear force or the nuclear in-medium effects. Given different categories of the effective Lagrangians, the residual interactions vary widely. For instance, the complexity of the residual interactions can considerably increase, when the meson-nucleon coupling strengths in the RMF and relativistic Hartree-Fock (RHF) models become density dependent \cite{Wang2020PRC101.064306}, in comparison to the models with constant coupling strengths. As results, it becomes rather challenging to incorporate them all into a single theoretical framework.

Nowadays, the finite amplitude method (FAM) provides an efficient approach to circumvent the calculations of the residual interactions \cite{Nakatsukasa2007PRC76.024318, Niksic2013PRC88.044327}. Another alternative approach is the time-dependent density functional theory (TD-DFT), a dynamic extension of the DFT. The TD-DFT is designed to describe the dynamical processes of many-body systems, including the giant resonances for small-amplitude collective oscillations and the reactions involving large-amplitude processes \cite{Nakatsukasa2016RMP88.045004}. For the small-amplitude collective motion, the linear density response of interacting systems can be rigorously formulated in the TD-DFT. In comparison to the RPA model, the TD-DFT owns a significant advantage of avoiding complicated calculations of the residual interactions, in which one only needs to calculate the time evolution of the one-body density matrix \cite{Ring1980NMP}. As a result, the computational complexity can be largely reduced when incorporating various effective Lagrangians into the TD-DFT framework.

The non-relativistic TD-DFT, which encompasses the Skyrme and Gogny types of interactions, has been developed extensively over time. In particular, the time-dependent Skyrme-Hartree-Fock model, based on the three-dimensional (3D) lattice space \cite{Maruhn2014CPC185.2195}, has been widely applied to numerous nuclear dynamical processes, such as the collective vibration \cite{Maruhn2005PRC71.064328, Schuetrumpf2016PRC93.054304}, multinucleon transfer process \cite{Simenel2010PRL105.192701}, fission \cite{Bulgac2016PRL116.122504, Tanimura2017PRL118.152501}, fusion \cite{Guo2018PLB782.401, Sun2023PRC107.064609} and cluster scattering \cite{Umar2010PRL104.212503}. In the case of the time-dependent Gogny-Hartree-Fock model and its corresponding time-dependent Gogny-Hartree-Fock-Bogoliubov model, the harmonic oscillator basis remains as a primary choice for calculations \cite{Hashimoto2012EPJA48.55, Hashimoto2013PRC88.034307, Hashimoto2016PRC94.014610}. The dynamical extension of the RMF model, known as the time-dependent RMF (TD-RMF), can be traced back to the early 1980s. At that time, the time-dependent versions of the Walecka model were employed to investigate the dynamics of colliding nuclear slabs \cite{Muller1981NPA372.459} and relativistic heavy ion collisions \cite{Cusson1985PRL55.2786}. Later, the TD-RMF model was employed extensively to investigate various GRs, including the ISGMR and the isovector giant monopole resonances (IVGMR) with spherical geometric symmetry \cite{Vretenar1997NPA621.853, Vretenar1997PRE56.6418, Vretenar1999NPA649.29c, Vretenar1999PRE60.308}, and the isovector giant dipole resonances (IVGDR), the isoscalar giant quadruple resonances (ISGQR) and isovector giant quadruple resonances (IVGQR) with axially deformed geometric symmetry \cite{Vretenar1995NPA581.679, Ring1996NPA598.107}. Moreover, the dynamics of Coulomb excitations of nuclei assuming axial symmetry have also been explored \cite{Vretenar1993PLB319.29}. In contrast to the non-relativistic TD-DFT, TD-RMF incorporates the time-odd fields naturally, which plays a pivotal role in describing the GRs. In a recent time, the success of solving the RMF equation in a 3D lattice space \cite{Ren2017PRC95.024313, Li2020PRC102.044307} led to the establishment of the TD-RMF model without any symmetry restrictions \cite{Ren2020PLB801.135194, Ren2020PRC102.044603}. More recently, the TD-RMF model was employed to investigate a multitude of processes, including the formation of linear-chain cluster states \cite{Ren2020PLB801.135194}, nuclear reactions \cite{Ren2020PRC102.044603, Zhang2024PRC109.024316, Zhang2024PRC109.024614}, chiral rotating \cite{Ren2022PRC105.L011301} and fission processes \cite{Ren2022PRC105.044313, Ren2022PRL128.172501, Li2023PRC107.014303, Li2024FP19.44201}.

In contrast to the RMF theory, that contains only the Hartree terms, the relativistic Hartree-Fock (RHF) theory explicitly incorporates the Fock terms of the meson-exchange diagram of nuclear force. It paves a natural way to incorporate the important ingredient of nuclear force. For instance, the tensor force is introduced automatically by the $\pi$ and $\rho$-tensor couplings, which contribute mainly via the Fock terms \cite{Long2008EPL82.12001, Jiang2015PRC91.034326}. Due to the Fock terms, significant improvements on describing the ground state properties have been achieved by the spherical and axially deformed RHF theory \cite{Long2006PLB640.150, Long2007PRC76.034314, Geng2020PRC101.064302}, and the extended relativistic Hartree-Fock-Bogoliubov (RHFB) theory \cite{Long2010PRC81.024308, Geng2022PRC105.034329}. Moreover, as demonstrated by the calculations of the RHF+RPA and RHFB+QRPA models, the Fock terms play an essential role in providing a self-consistent description of the spin-isospin excitation of nuclei \cite{Liang2008PRL101.122502, Liang2009PRC79.064316, Liang2012PRC85.064302, Niu2013PLB723.172, Niu2017PRC95.044301, Wang2020PRC101.064306}. Thus, it is of interest to investigate the impact of Fock terms on both small-amplitude collective oscillations and large-amplitude processes. This motivates us to develop the dynamical extension of the RHF theory, namely the time-dependent RHF (TD-RHF) model.

In this work, as a preliminary step, we will establish the spherical TD-RHF model by focusing on the nuclear ISGMR mode, and the Dirac spinors are expanded on the spherical Dirac Woods-Saxon (DWS) basis \cite{Zhou2003PRC68.034323}. The paper is organized as follows. In Sec. \ref{sec:GF}, a brief overview of the comprehensive theoretical framework is provided. This includes a description of the time-dependent RHF equations solved utilizing the spherical DWS basis, as well as the numerical details pertaining to the simulation of the nuclear ISGMR mode. Section \ref{sec:RD} presents the numerical tests and primary applications of the ISGMR of $^{208}$Pb. Finally, a summary is provided in Sec. \ref{sec:sum}.

\section{General Formalism}\label{sec:GF}

This section provides a concise overview of the general formalism of the relativistic Hartree-Fock (RHF) theory, followed by a introduction of the TD-RHF model utilizing the spherical DWS basis. Furthermore, the numerical details involved in simulating the ISGMR mode using the TD-RHF framework will be introduced.

\subsection{Static and Dynamic RHF model}

Based on the meson-exchange diagram of nuclear force, the RHF Lagrangian encompasses the degrees of freedom associated with nucleon ($\psi$), isoscalar scalar $\sigma$-meson, isoscalar vector $\omega$-meson, isovector vector $\rho$-meson, isovector pseudoscalar $\pi$-meson and photon ($A$) fields. Among the selected degrees of freedom, the isoscalar $\sigma$- and $\omega$-meson fields dominate strong attractive and repulsive nucleon-nucleon interactions, respectively, the isovector $\rho$- and $\pi$-meson fields are introduced to describe the isospin-related properties, and the photon field for the electra-magnetic interactions between protons.

Following the Legendre transformation, one can obtain the Hamiltonian of nuclear systems. Substituting the field equations of nucleons, mesons and photons, the Hamiltonian operator can be generally expressed as,
\begin{align}\label{eq:Hamiltonian}
    \hat H = \hat T + \sum_\phi \hat V_\phi,
\end{align}
where the kinetic energy ($\hat T$) and potential energy ($\hat V_\phi$) terms read as,
\begin{align}
    \hat T=& \int d\svec x \bar\psi(\svec x) \left(-i\svec\gamma\cdot\svec\nabla + M\right) \psi(\svec x), \\
    \hat V_\phi =& \ff2 \int d\svec xd\svec x' \bar\psi(\svec x)\bar\psi(\svec x') \Gamma_\phi D_\phi \psi(\svec x)\psi(\svec x'),
\end{align}
with $M$ for nucleon mass. In the potential energy terms $V_\phi$, the symbols $\Gamma_\phi$ represent the interaction vertices associated with various coupling channels, including the Lorentz scalar ($\sigma$-S), vector ($\omega$-V, $\rho$-V, $A$-V), tensor ($\rho$-T), vector-tensor ($\rho$-VT) and pseudo-vector ($\pi$-PV) couplings, and $D_\phi$ denotes the propagators of meson and photon fields. Further details are referred to Refs. \cite{Geng2020PRC101.064302, Geng2022PRC105.034329, Long2022CTP74.097301}.

In this work, the relativistic Hartree-Fock approach is imposed for both static and dynamic case. Consistent with that, the contributions from the negative energy states are neglected, namely the no-sea approximation, which is usually adopted for the static calculations. For the dynamic case, here we focus on the ISGMR mode, a small-amplitude collection motion. Under the TD-RHF framework, it still remains a reasonable choice to ignore the contributions of Dirac sea to the one-body nucleonic density. Thus, the nucleon field operator $\psi$ in the Hamiltonian can be quantized as
\begin{align}\label{eq:SE}
    \psi(x) = \sum_i \psi_i(\svec x) e^{-i\varepsilon_i t} c_i,
\end{align}
where the annihilation and creation operators $c_i$ and $c_i^\dag$ are defined by the positive energy solutions of the Dirac equation, $\varepsilon_i$ is the single-particle energy ($\varepsilon_i>0$), and $\psi_i(\svec x)$ is the Dirac spinor of state $i$. Following the quantization (\ref{eq:SE}) of nucleon field $\psi$, the initial Hartree-Fock ground state $\lrlc{\text{HF}}$ can be deduced, and the expectation of the Hamiltonian (\ref{eq:Hamiltonian}) with respect to $\lrlc{\text{HF}}$ gives the total energy functional $E$ of nuclear system,
\begin{align}\label{eq:ENE}
    \lrlc{\text{HF}} = & \prod_{i=1}^A c_i^\dag \lrlc{-}, &E = & \lrcl{\text{HF}} \hat H \lrlc{\text{HF}},
\end{align}
where $A$ is nuclear mass number and $\lrlc{-}$ represents the vacuum state. It shall be emphasized that the expectation of the two-body interaction parts, namely $V_\phi$, contains two types of contributions, the direct (Hartree) and exchange (Fock) terms. Due to the non-local Fock terms, an integro-differential Dirac equation of nucleons is obtained from the variation of the energy functional (\ref{eq:ENE}),
\begin{align}\label{eq:RHF}
    \int d\svec r \hat h(\svec x,\svec x') \psi_i(\svec x') = e_i \psi_i(\svec x),
\end{align}
where $e_i$ is the single-particle energy of the state $i$, and $\hat h$ represents the single-particle Hamiltonian. Specifically, $h$ consists of the kinetic terms $\hat h^{\text{kin}}$, the local mean potential term $\hat h^D$ and the non-local potential term $\hat h^E$,
\begin{align}
    \hat h^{\text{kin}}(\svec x,\svec x') = & \Big[ \svec\alpha\cdot\svec p + \gamma^0 M \Big] \delta(\svec x-\svec x'), \\
    \hat h^{D}(\svec x,\svec x') =& \Big[\Sigma_T(\svec x)\gamma_5 + \Sigma_0(\svec x) + \gamma^0 \Sigma_S(\svec x)\Big] \delta(\svec x-\svec x'), \\
    \hat h^{E}(\svec x,\svec x') =& \begin{pmatrix} Y_G(\svec x,\svec x') & Y_F(\svec x,\svec x') \\[0.5em] X_G(\svec x,\svec x') & X_F(\svec x,\svec x') \end{pmatrix}.
\end{align}
In the above expressions, $\Sigma_S$, $\Sigma_0$ and $\Sigma_T$ denote the scalar potential, the time component of the vector potential and the tensor potential, respectively, and the terms $X_G$, $X_F$, $Y_G$, and $Y_F$ are the non-local potentials contributed by the Fock terms.

For the dynamic case, the time-dependent (TD) RHF equation can be derived through a similar routine. According to the standard procedure described in Ref. \cite{Ring1980NMP}, the time-dependent many-body problem can be reduced to a time-dependent one-body equation. This reduction allows us to obtain the evolution of the single-nucleon wave function $\psi_i$, which fulfills the TD-RHF equation as,
\begin{align}\label{eq:TDRHF}
    i\frac{\partial}{\partial t} \psi_i(t,\svec x) = \int d\svec r'\hat h(\svec x,\svec x',t) \psi_i(t,\svec x'),
\end{align}
where the index $i$ represents the single-particle states. The time-dependent Hamiltonian $\hat h(\svec x,\svec x',t)$ is fully determined by the time-dependent local density, local currents and non-local density. Therefore, under the adiabatic approximation, the time-dependent single-particle Hamiltonian $\hat h(\svec x,\svec x',t)$ can be obtained using the wave function $\psi_i(t,\svec x)$ at a given time $t$.

It should be noticed that in the static RHF theory, the spatial component of the four-dimension current $j_\mu$, namely $\svec j$, vanishes due to the time-reversal symmetry. However, in the time-dependent case, the initial excitation breaks the time-reversal symmetry at a particular time step. Consequently, the current $\svec j$ exerts an influence on the RHF mean field during the time evolution process.

\subsection{Spherical TD-RHF model with the DWS basis}

Regarding the numerical difficulties introduced by the Fock terms, we impose the spherical symmetry to investigate the nuclear ISGMR initially, and solve the TD-RHF equation (\ref{eq:TDRHF}) by expanding the nucleon wave functions on the Dirac Woods-Saxon (DWS) basis \cite{Zhou2003PRC68.034323}. Under the spherical symmetry, the complete set of good quantum numbers comprises the principle one $n$, the total angular momentum $j$ and its projection $m$, as well as the parity $\pi=(-1)^l$ ($l$ represents the orbital angular momentum). Introducing the quantum number $\kappa$, i.e., $\kappa=\pm(j+1/2)$ with $j=l\mp1/2$, the wave function of the states in the spherical DWS base reads as,
\begin{align}
    \psi_{a\kappa m} (\svec x) = \frac{1}{r} \begin{pmatrix} G_{a\kappa}(r) \Omega_{\kappa m} (\vartheta,\varphi) \\[0.5em] iF_{a\kappa}(r) \Omega_{-\kappa m} (\vartheta,\varphi) \end{pmatrix},
\end{align}
with $\svec x = r \svec e_r(\vartheta,\varphi)$. In order to avoid confusion, we use the index $a$ and $n$ to denote the DWS basis states and the time-dependent ones of nucleus, respectively. Thus, in terms of the DWS base, the time-dependent wave function $\psi_{n\kappa m}(t,\svec r)$ can be expanded as,
\begin{align}\label{eq:expansion}
    \psi_{n\kappa m}(t, \svec x) = \sum_{a} C_{na,\kappa}(t) \psi_{a\kappa m}(\svec x).
\end{align}
It should be noticed that the expansion coefficients $C_{na,\kappa}(t)$ carry all the time-evolution information of the system, regarding the completeness of the DWS base. Consequently, the TD-RHF equation (\ref{eq:TDRHF}) in coordinate space can be transformed into the DWS basis space,
\begin{align}\label{eq:C-T}
    i\frac{\partial}{\partial t} C_{na,\kappa}(t) = \sum_{b} H_{ab,\kappa}(t) C_{nb,\kappa}(t),
\end{align}
where both $a$ and $b$ are the index of the DWS basis states. The Hamiltonian matrix element $H_{ab,\kappa}$ for the $\kappa$-block can be obtained as,
\begin{align}
   \big[ H_{\kappa}(t)\big]_{ab} \equiv & \int d\svec xd\svec x' \sum_m\psi_{a\kappa m}^\dag(\svec x) \hat h(\svec x,\svec x',t) \psi_{b\kappa m}(\svec x').
\end{align}
It is worth to noting that the use of DWS basis largely reduces the numerical calculations for solving the TD-RHF equation, without loosing essential physics.

In order to solve Eq. (\ref{eq:C-T}), the time-evolution matrix $U_{ab,\kappa}(t,t_0)$, which describes the evolution of the single-particle state from the initial time $t_0$ to the final time $t$, is introduced. Thus, the expansion coefficient at any time $t$ can be obtained as,
\begin{align}\label{eq:C}
    C_{na,\kappa}(t) = \sum_b U_{ab,\kappa}(t,t_0) C_{nb,\kappa}(t_0).
\end{align}
Following the principles of perturbation theory \cite{Das2021QFT}, the time-evolution matrix element $U_{ab,\kappa}$ can be formally expressed as,
\begin{align}
    U_{ab,\kappa}(t,t_0) = \hat{\cals{T}} \Big[\exp\Big(-i\int_{t_0}^t H_{\kappa}(t')dt'\Big)\Big]_{ab},
\end{align}
where $\hat{\cals T}$ represents the time-ordering operation. Substituting it into Eq. (\ref{eq:C}), one obtains the time-evolution of the expansion coefficient. This allows us to calculate the evolution of expansion coefficient over time with given initial condition.

\subsection{Numerical Details}

The predictor-corrector strategy \cite{Maruhn2014CPC185.2195} is adopted for the numerical implementation of the formal solution (\ref{eq:C}). The evolution time is discretized into a series of small time step $\Delta t$. For a given time interval $[t,t+\Delta t]$, the single Hamiltonian matrix in Eq. (\ref{eq:C}) is approximated by its value at the mid-time $H_{ab,\kappa}(t_m = t+\Delta t/2)$. Consequently, the evolution of the expansion coefficient from $t$ to $t+\Delta t$ is approximated as,
\begin{align}\label{eq:approx}
    C_{na,\kappa}(t+\Delta t) \approx  &\sum_b \big[ e^{ -iH_{\kappa}(t_m)\Delta t} \big]_{ab} C_{nb,\kappa}(t),
\end{align}
where the Hamiltonian matrix $H_{ab}^i(t_m)$ is determined through the predictor-corrector strategy. Firstly, the local and nonlocal densities/currents at time $t+\Delta t$, uniformly identified as $\tilde\rho(t+\Delta t)$, are calculated by using the intermediate expansion coefficient $\tilde C_{i,a}(t+\Delta t)$,
\begin{align}
    \tilde C_{na,\kappa}(t+\Delta t) =  &\sum_b \big[e^{ -i H_\kappa(t)\Delta t} \big]_{ab} C_{nb,\kappa}(t).
\end{align}
Secondly, the Hamiltonian matrix element $H_{ab,\kappa}(t_m)$ in Eq. (\ref{eq:approx}) is approximately constructed using the average $\big[\rho(t) + \tilde\rho(t+\Delta t)\big]/2$ as the inputs. Further, the obtained Hamiltonian matrix $H_{ab,\kappa}(t_m)$ is substituted into Eq. (\ref{eq:approx}) to get the expansion coefficient $C_{na,\kappa}(t+\Delta t)$.

Moreover, the exponential function of the Hamiltonian operator is evaluated by using a Taylor expansion up to order $m$ as the following form,
\begin{align}\label{eq:m}
    \big[ e^{-i\Delta t H_\kappa }\big]_{ab} \approx \sum_{n=0}^m \frac{\big( -i\Delta t\big)^n}{n!} \big[ H_\kappa^n\big]_{ab}.
\end{align}
It is important to note that the truncation of the Taylor expansion must be tested to maintain good unitarity of $\exp\left(-iH\Delta t\right)$ and energy conservation. Therefore, careful checks on the conservation of particle number and energy are necessary to ensure the quality of the time evolution. Section \ref{sec:RD} presents the numerical check on the time step $\Delta t$ and truncation $m$ in detail. If not particularly specified, the values of $\Delta t = 0.1$ fm/c and $m=4$ will be adopted in the following calculations.

To excite ISGMR, an appropriate external perturbation must be applied. Typically, the isoscalar giant monopole excitation operator $\hat F_{\text{IS}}$ is used both to create and measure the oscillations. This operator is defined as,
\begin{align}
    \hat F_{\text{IS}} = r^2,
\end{align}
where $r$ represents the radial distance from the center of a nucleus. The initial condition for the present calculation is assumed as an attractive type of perturbation. The static RHF solutions $C^{\text{HF}}$ are transformed into the initial matrices $C^{0}$ through an instantaneous boost, according to the following relation \cite{Pardi2013PRC87.014330},
\begin{align}\label{eq:initial}
    C^{0}_{na,\kappa} = \sum_b \exp\left(-ik F_{ab,\kappa} \right) C_{nb,\kappa}^{\text{HF}}.
\end{align}
Here $F_{ab,\kappa}$ stands for the matrix elements of the isoscalar giant monopole excitation operator $F$ for block $\kappa$,
\begin{align}
    & F_{ab,\kappa} \equiv \sum_m \int d\svec x \psi_{a\kappa m}^\dag(\svec x) r^2 \psi_{b\kappa m}(\svec x).
\end{align}
The parameter $k$ in Eq. (\ref{eq:initial}) controls the strength of the initial boost and is chosen to ensure the linearity in the course of the time integration. In this work, the value of parameter $k$ is set as $0.01$ fm$^{-2}$.

The initial condition (\ref{eq:initial}) at t = 0 serves as the starting point for a dynamic evolution of the system, and  the $F$-signal can be measured over time. This signal is given by,
\begin{align}
    F(t) = \sum_{n\kappa m} \int d\svec x \psi_{n\kappa m}^\dag(t,\svec x) \hat F \psi_{n\kappa m}(t,\svec x),
\end{align}
where $\psi_{n\kappa m}(t,\svec x)$ represents the time-dependent single-particle wave functions. In practice, we are interested in the deviation of the $F$-signal from its initial value at $t=0$,
\begin{align}\label{eq:F-signal}
    F(t) \rightarrow F(t) - F(t=0).
\end{align}
By subtracting the initial value, we focus on the changes in the $F$-signal that are due to the applied perturbation. The strength function, which characterizes the response of the system to the external perturbation, can be obtained from the Fourier transform of the $F$-signal deviation \cite{Nakatsukasa2005PRC71.024301},
\begin{align}\label{eq:Strength Fun}
    S(\omega) = -\frac{1}{\pi k} \text{Im} \int dt F(t) e^{i\omega t},
\end{align}
where Im stands for the imaginary part of the expression, and $\omega$ represents the frequency of the response. For the isoscalar giant monopole excitation, the strength function $S(\omega)$ can give information about the resonant frequencies and their corresponding strengths in the response of the system.



\section{Results and Discussion} \label{sec:RD}

\subsection{Numerical Check}

In the static and dynamic calculations, the spherical DWS base is utilized. The box size is set to be 24 fm, with a mesh spacing of 0.1 fm. The energy cutoffs, denoted as $E_{\pm}^C$, represent the positive $(+)$ and negative $(-)$ energy cutoffs in the spherical DWS base. Specifically, the states with positive (negative) energies $E$, that satisfy the conditions $E - M < E_{+}^C$ ($E + M > E_-^C$) are considered in the expansion (\ref{eq:expansion}). Further details on the energy cutoffs in the DWS basis can be found in Ref. \cite{Geng2022PRC105.034329}. After careful tests, the energy cutoffs $E_+^C$ and $E_-^C$ are set to be $+400$ MeV and $-100$ MeV, respectively, for both static and dynamic calculations.

In order to ensure an accurate description of nuclear ISGMR and to maintain the quality of the time evolution, it is crucial to perform a convergence check for the total energy and particle number, with respect to the time step ($\Delta t$) in Eq.~(\ref{eq:approx}) and the Taylor expansion order ($m$) in Eq.~(\ref{eq:m}). In this section, the conservation of the total energy excluding the center-of-mass correction, and the particle number are checked, as well as the time reversal invariance for the ISGMR mode of $^{208}$Pb. 

\begin{figure}[htbp]
  \centering\setlength{\abovecaptionskip}{0.5em}
  \includegraphics[width=0.98\linewidth]{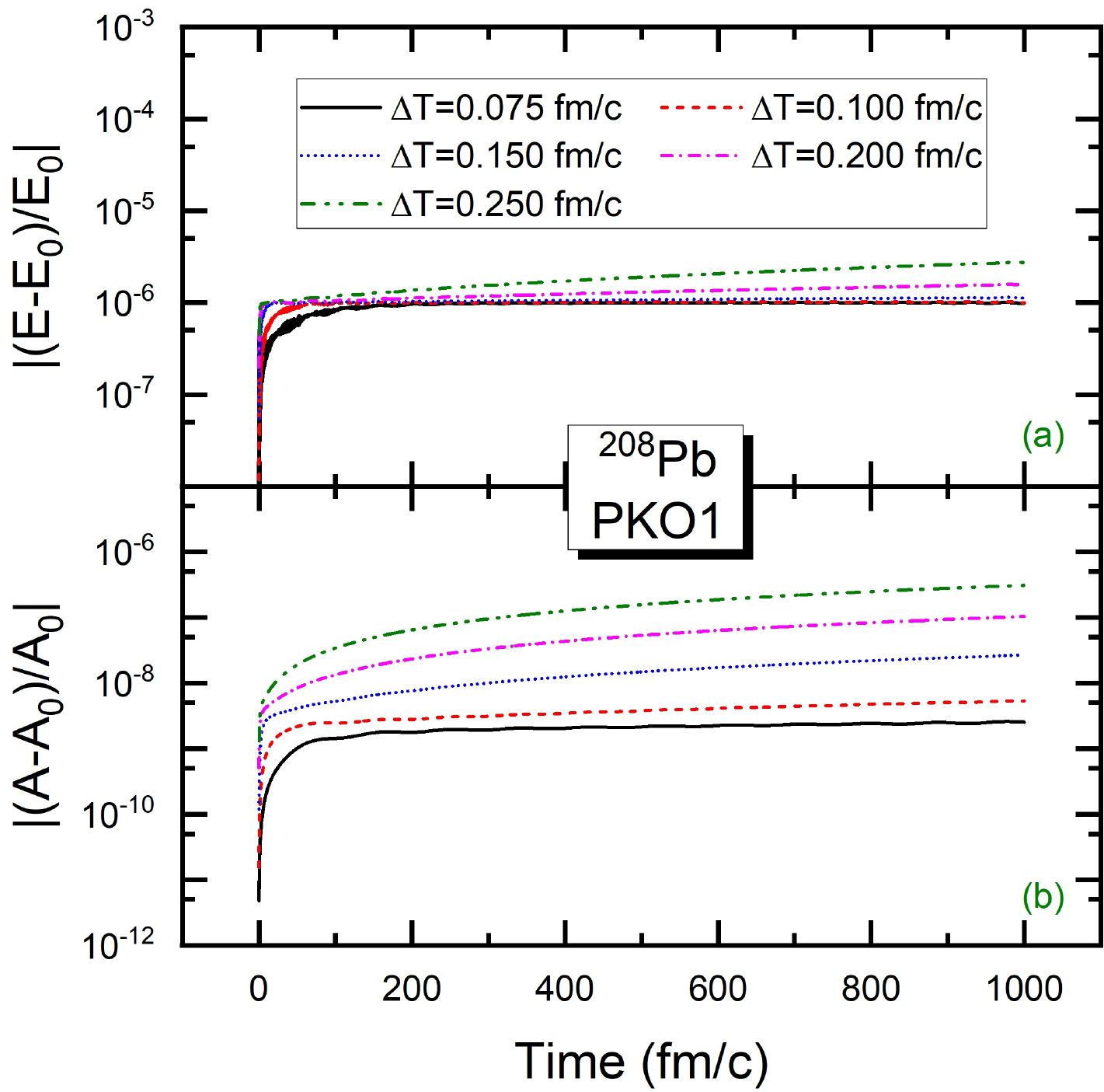}\\
  \caption{(Color Online) (a) The relative energy deviation $\left| \left( E-E_{0}\right)/E_{0} \right|$ (panel a) and the relative particle number deviation $\left| \left( A - A_{0} \right)/A_{0} \right|$ (b) with respect to the initial Energy $E_{0}$ and initial particle number $A_{0}$. This picture shows the results with $\Delta t$ = 0.075 fm/c, 0.1 fm/c, 0.15 fm/c, 0.2 fm/c and 0.25 fm/c with $m=4$.}\label{Fig:T}
\end{figure}

Firstly, the convergence of the time step $\Delta t$ is verified. Figure \ref{Fig:T} presents the time evolutions of the relative energy deviation $\left| [E-E_{0}]/E_{0} \right|$, and relative particle number deviation $\left| [A-A_{0}]/A_{0} \right|$ for various values of $\Delta t$. The selected time steps $\Delta t$ are $0.075, 0.1, 0.15, 0.2$ and $0.25$ fm/c, with the cut-off value $m=4$ in the Taylor expansion (\ref{eq:m}). According to Fig. \ref{Fig:T} (a), it is evident that the relative energy deviations are approximately $10^{-6}$ for all the chosen values of the time step $\Delta t$. As the time step $\Delta t$ decreases, the relative energy deviations exhibit a more uniform and stable behaviour over longer evolution times. It indicates that the calculations are converging with respect to the time step. The conservation of particle number $A$ is more robust than the total energy. As shown in Fig. \ref{Fig:T} (b), it can be observed that as the time step is reduced, the relative particle number deviations are rather stable as well. Notably, for a time step of $\Delta t$ = 0.1 fm/c, the relative particle number deviations are approximately $10^{-8}$. Such a precision can be attributed to the fact that the approximations in Eq. (\ref{eq:approx}) become more accurate for smaller values of $\Delta t$. Consequently, the convergence analysis demonstrates that a time step of $\Delta t=0.1$ fm/c, combined with a Taylor expansion order of $m=4$, is accurate enough for the calculation of the ISGMR in $^{208}$Pb.

\begin{figure}[htbp]
  \centering\setlength{\abovecaptionskip}{0.5em}
  \includegraphics[width=0.98\linewidth]{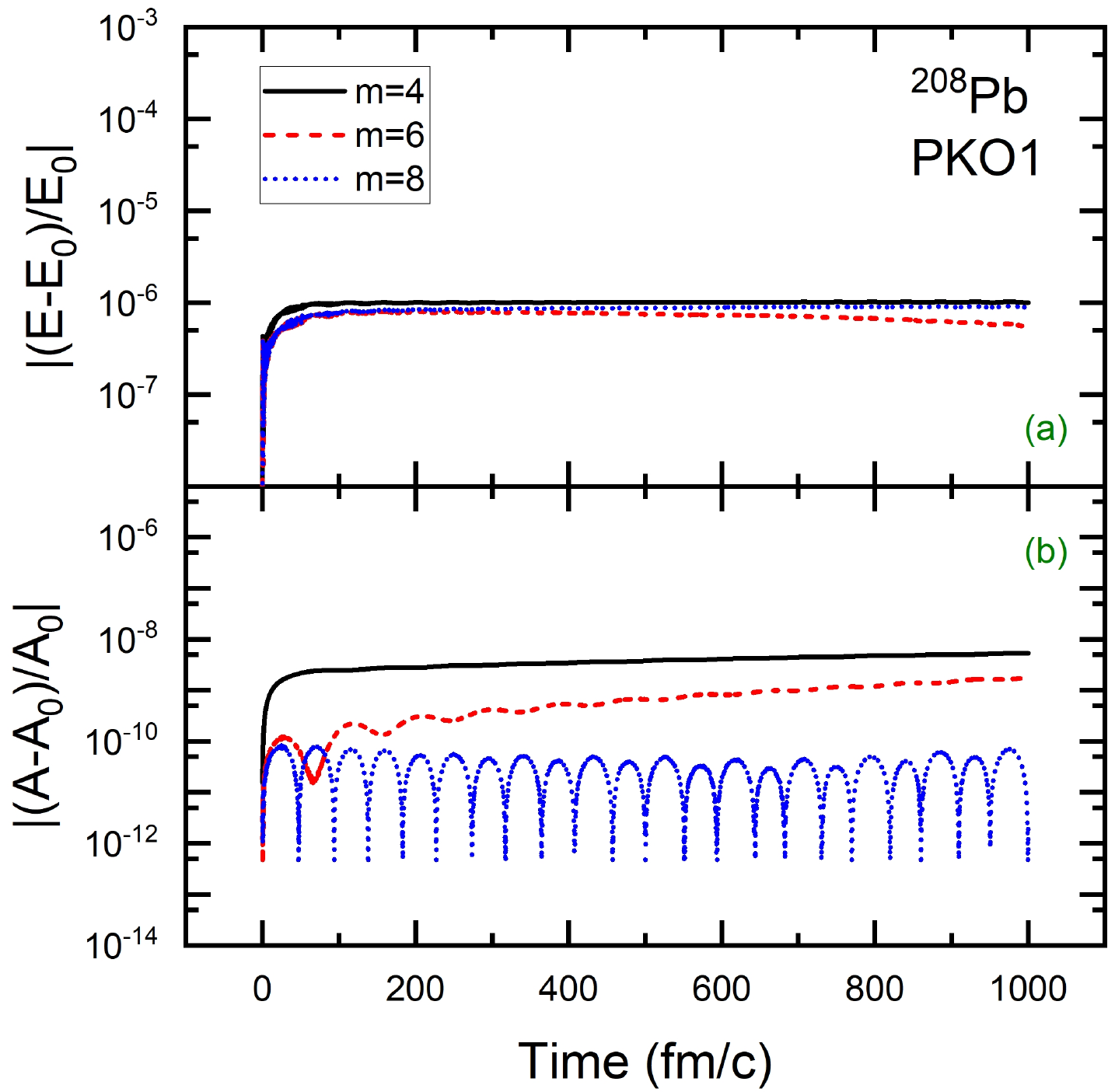}\\
  \caption{(Color Online) Be similar with Fig. (\ref{Fig:DT}), but shows the results with $m=4,6$ and $8$ with $\Delta t=0.1$ fm/c. }\label{Fig:m}
\end{figure}

To illustrate further, we verify the convergence with respect to the cut-off of $m$ in the Taylor expansion (\ref{eq:m}). Figures \ref{Fig:m} (a) and (b) present the time evolution of the relative energy deviation $\left| [E-E_{0}]/E_{0} \right|$ and relative particle number deviation $\left| [A-A_{0}]/A_{0} \right|$ with different $m$ values, respectively. It is clear that with the $m$ value increasing, the conservation of total energy and particle number improves. This observation is consistent with our understanding that the cut-off value of $m$ directly impacts the conservation properties, as previously discussed. For $m=4$, the relative total energy deviation is approximately $10^{-6}$ and the relative particle number deviation is approximately $10^{-8}$, which promises the precision in calculating the ISGMR of $^{208}$Pb. Further increasing the $m$ value, the conservation of the total energy and particle number can be even more robust. Notably, even for $m=8$, rather small relative deviation of the particle number around $10^{-11}$ can ensure the particle number conservation, despite a regular oscillation with the time evolution. The convergence check indicates that the cut-off value of $m=4$ provides a satisfied precision for the calculation of the ISGMR in $^{208}$Pb, and larger $m$ values may further improve the conservations but not quite necessary.

\begin{figure}[htbp]
  \centering\setlength{\abovecaptionskip}{0.5em}
  \includegraphics[width=0.98\linewidth]{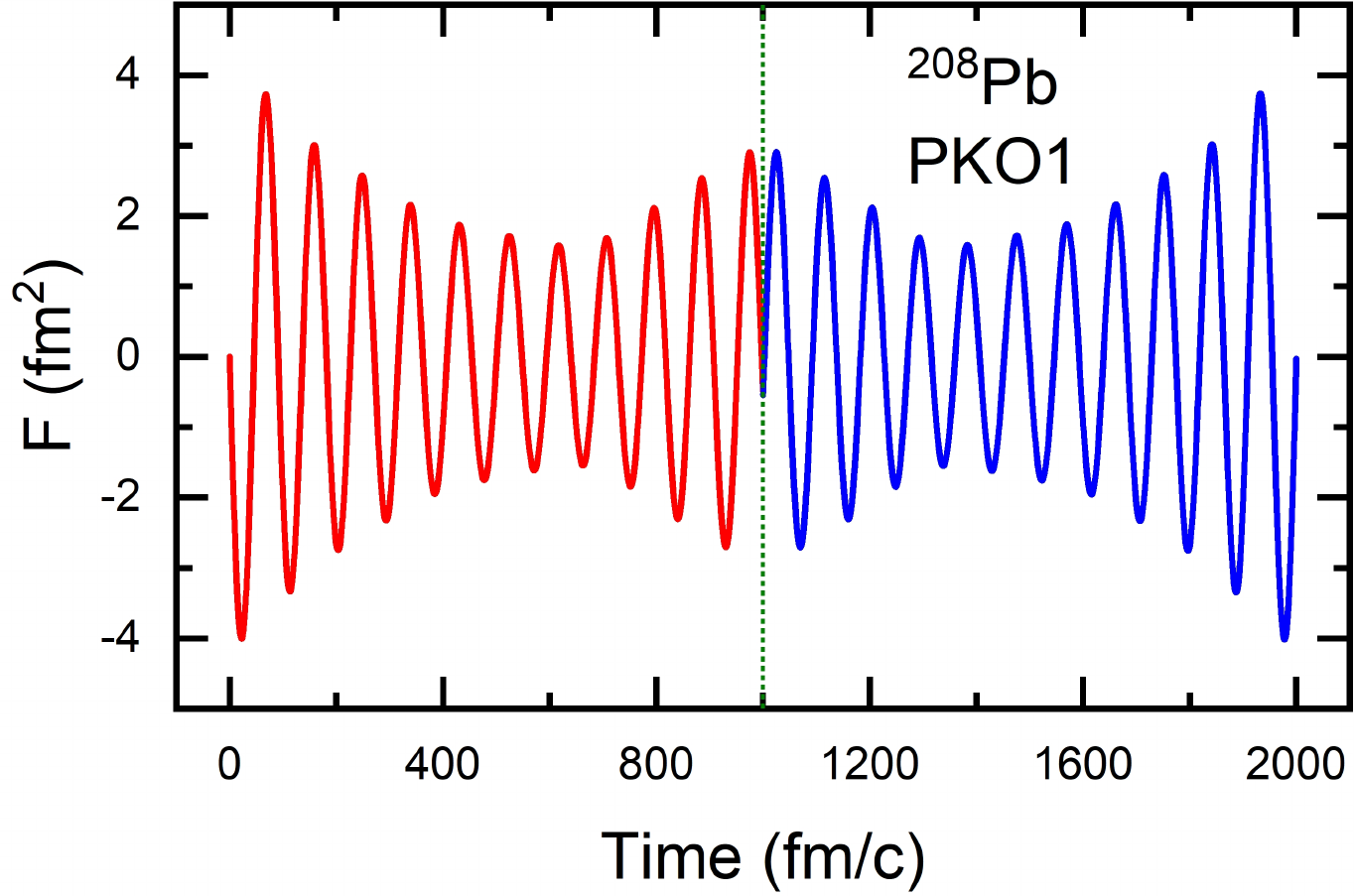}
  \caption{(Color Online) Time evolution of ISGMR signal $F$. The single particle wave functions at time $t=1000$fm/c are replaced by their time-reversal conjugates.}\label{Fig:inver}
\end{figure}

Furthermore, a crucial test for the TD-RHF framework is provided by the time-reversal invariance, which is a fundamental property that ensures the microscopic reversibility of the entire system. To demonstrate this property in the $^{208}$Pb ISGMR, the single-particle wave functions $\psi_i(r,t)$ at time $t$ = 1000 fm/c are replaced by their time-reversal conjugates $\hat T \psi_i(r,t)$. Theoretically, if time-reversal invariance is preserved, the system should revert to its initial state as time progresses. Figure \ref{Fig:inver} presents the time evolution of the ISGMR signal $F(t)$. It is obvious that after substituting $\psi_i(r,t)$ with $\hat T\psi_i(r,t)$ at 1000fm/c, $F(t)$ indeed evolves back precisely to its original state. This observation provides unambiguous evidence that the time-reversal invariance is satisfied in current TD-RHF calculations. The preservation of time-reversal invariance ensures the consistency and reliability of the calculations.

Based on the numerical checking, it can be concluded that the total energy, particle number and time-reversal invariance are conserved with remarkable precision in the current TD-RHF calculations with time step of $\Delta t$ = 0.1 fm/c and cut-off value for Taylor expansion of $m=4$. These high qualities of conservation ensure the precision and reliability of our results and validate the TD-RHF approach for describing the nuclear ISGMR mode.

\subsection{ISGMR of $^{208}$Pb}

After a careful numerical examination, the ISGMR mode is explored utilizing the newly developed TD-RHF model. In this work, we present the results for the double-magic nucleus $^{208}$Pb, because $^{208}$Pb is the obvious benchmark and there are lots of experimental results for ISGMR model of $^{208}$Pb. According to the experimental data in 2013, the peak position of the strength function for $^{208}$Pb is 13.7$\pm$0.1 MeV, and the constrained energy is reported to be $13.5\pm0.1$ MeV~\cite{Patel2013PLB726.178}. The value of constrained energy can be extracted from the strength function. The $k$-th moments of the strength function $S(\omega)$ are defined as,
\begin{align}\label{eq:centroid}
    m_k = \int_{0}^\infty S(\omega) \omega^k d\omega.
\end{align}
Subsequently, the constrained energy is calculated as the moment ratios $\sqrt{m_1/m_{-1}}$.

\begin{figure}[htbp]
  \centering\setlength{\abovecaptionskip}{0.5em}
  \includegraphics[width=0.98\linewidth]{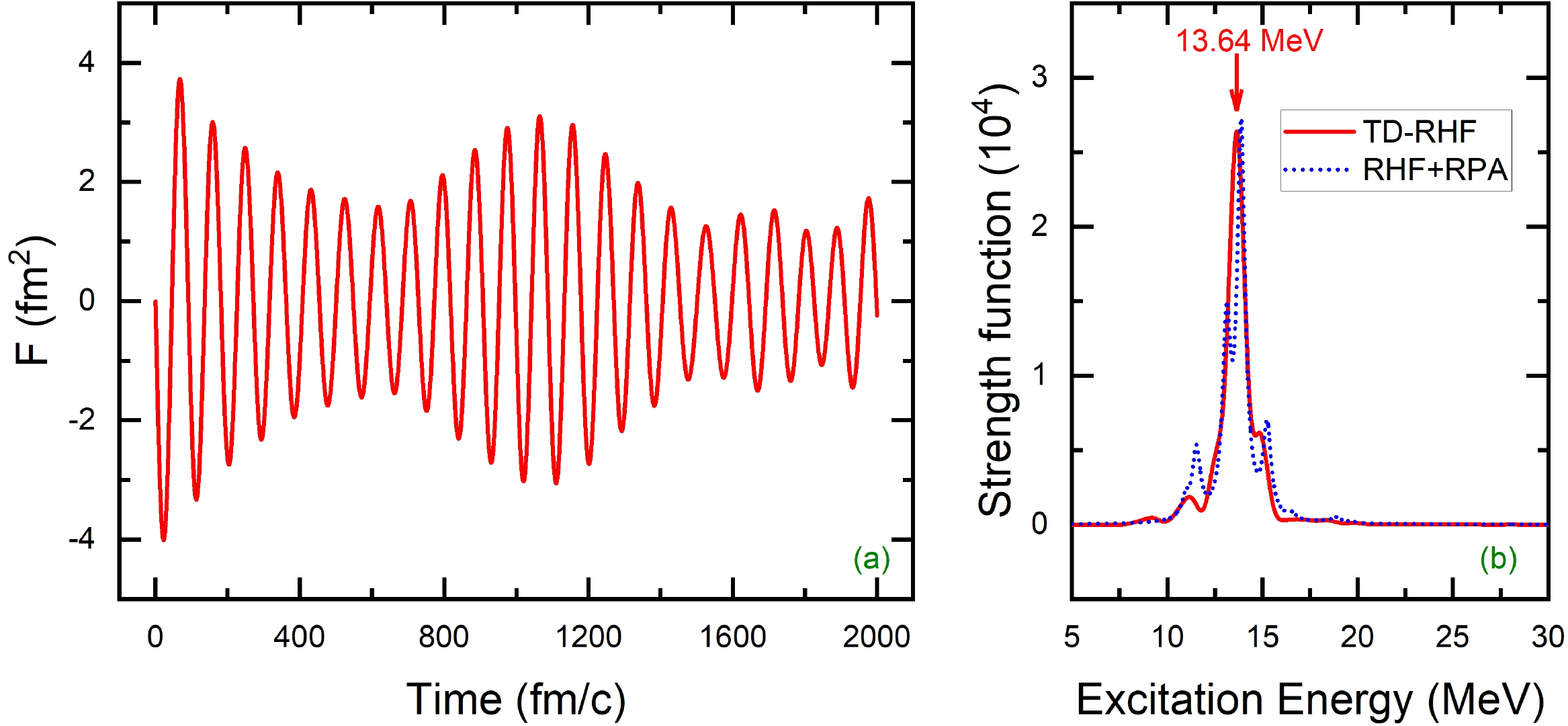}\\
  \caption{(Color Online) Time-dependent $F$-signal isoscalar (panel a) and the corresponding strength spectra for $^{208}$Pb with the peak position (panel b). The blue dot line in panel (b) denotes the results calculated by RHF+RPA model with the width $\Gamma$ is taken to be 0.5 MeV in Lorentzian function. The results are calculated using the effective Lagrangian PKO1. }\label{Fig:DT}
\end{figure}

Firstly, the TD-RHF calculation is performed utilizing the RHF effective Lagrangian PKO1. Figures \ref{Fig:DT} (a), (b) present the $F$-signal defined in Eq. (\ref{eq:F-signal}) and the corresponding strength function of ISGMR as shown in Eq. (\ref{eq:Strength Fun}), respectively. As illustrated in Fig. \ref{Fig:DT} (a), the $F$-signal calculated by PKO1 effectively reproduces the regular oscillations that are characteristic of the ISGMR mode. The selected final evolution time, $T_{\text{final}}=$ 2000 fm/c, determines the numerical resolution in the frequency domain of approximately $\Delta E = \pi \hbar /T_{\text{final}} \approx 0.31$ MeV \cite{Reinhard2006PRE73.036709}. In accordance with expectations for a heavy nucleus, there is minimal spectral fragmentation in the isoscalar channel, with a single mode dominating the excitation energy. Consequently, a distinct peak with excitation energy 13.64 MeV is observed in the strength function, as shown in Fig. \ref{Fig:DT} (b), which is consistent with the experimental data, 13.7$\pm$0.1 MeV ~\cite{Patel2013PLB726.178}.

As a comparison, Fig.~\ref{Fig:DT} (b) presents the strength function calculated by the RHF+RPA model, revealing a peak position of strength function of about 13.84 MeV. Although there is a deviation of approximately 0.2 MeV for peak position between the TD-RHF and RHF+RPA models, the areas of strength function from 5 MeV to 30 MeV read as 39980.06 and 39945.38 for RHF+RPA and TDRHF models, respectively. Hence the deviation of the peak position is mainly due to the weak splitting in the strength function given by the RHF+RPA model, as illustrated in Fig.~\ref{Fig:DT} (b). Furthermore, according to the strength function in Fig.~\ref{Fig:DT} (b) and the definition of the moments for strength function in Eq.~(\ref{eq:centroid}), the constrained energy of ISGMR for $^{208}$Pb calculated by TD-RHF model with PKO1 is 13.61 MeV with an integrating range 9.5 MeV $\sim$ 19.5 MeV. The constrained energy calculated using the RHF+RPA model with PKO1 is 13.62 MeV, which provides further corroboration of the reliability of the newly developed TD-RHF model. Both of the obtained values, $13.61$ MeV and $13.62$ MeV, are close to the experimental data $13.5\pm0.1$ MeV, thereby validating the reliability of the TD-RHF model with PKO1.

\begin{figure}[htbp]
  \centering\setlength{\abovecaptionskip}{0.5em}
  \includegraphics[width=0.98\linewidth]{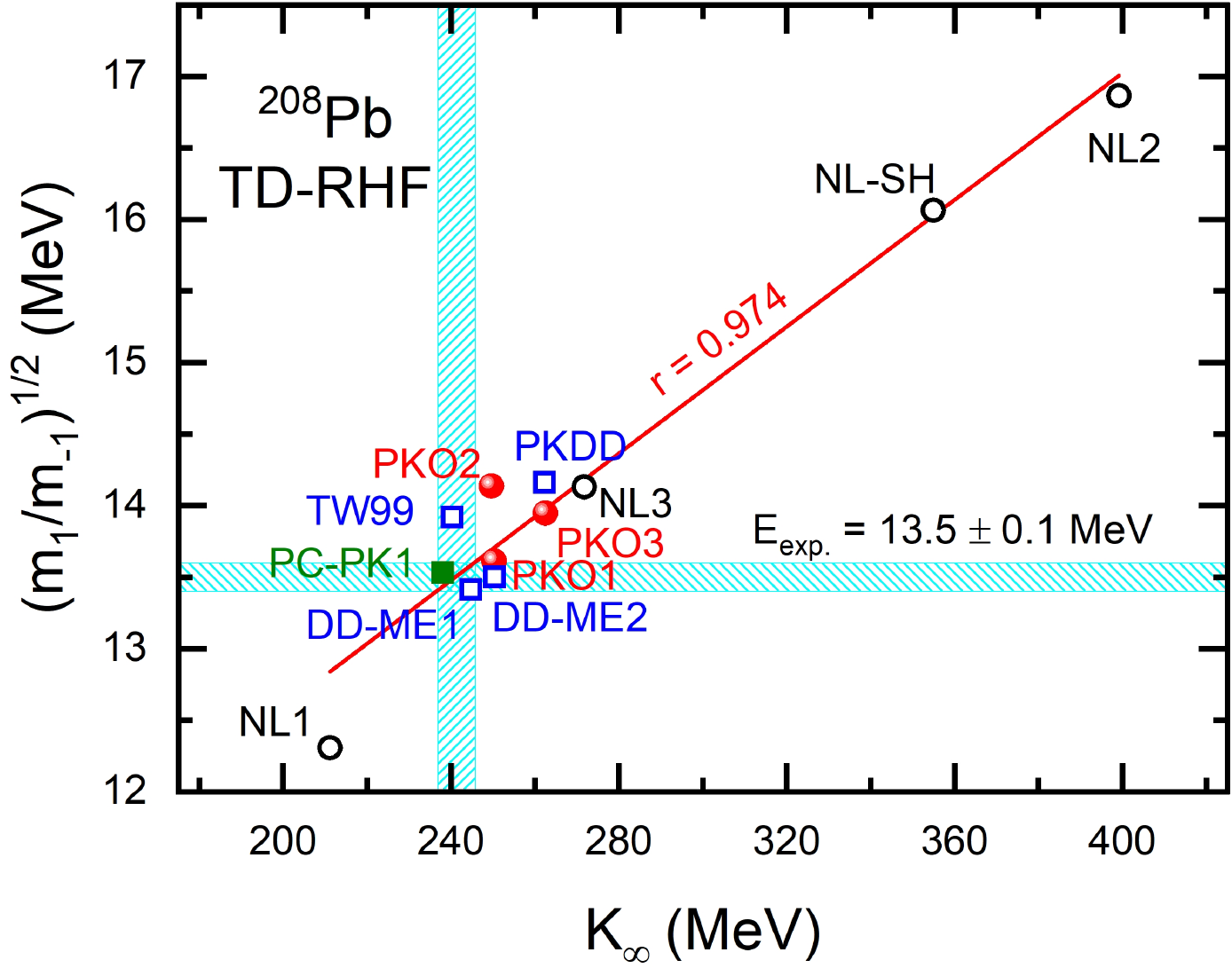}\\
  \caption{(Color Online) The constrained energy $\sqrt{m_1/m_{-1}}$ of ISGMR in $^{208}$Pb vs nuclear incompressibility $K_\infty$ (MeV). The results are calculated by RHF effective Lagrangian PKO$i$($i$=1,2,3) \cite{Long2006PLB640.150, Long2008EPL82.12001}, RMF effective Lagrangian DD-ME1 \cite{Niksic2002PRC66.024306}, DD-ME2 \cite{Lalazissis2005PRC71.024312}, PKDD \cite{Long2004PRC69.034319}, TW99 \cite{Typel1999NPA656.331}, NL$i$($i$=1,2,3) \cite{Reinhard1986ZPA323.13, Lee1986PRL57.2916, Lalazissis1997PRC55.540}, NLSH \cite{Sharma1993PLB312.377} and PC-PK1 \cite{Zhao2010PRC82.054319}. The red line is the linear fitting between the constrained energy of ISGMR and $K_\infty$ with Pearson's r 0.979, and the black line is the results calculated by NL$i$ ($i$=1,2,3) and NL-SH. The experimental data of constrained energy for ISGMR in $^{208}$Pb is taken from \cite{Patel2013PLB726.178} }\label{Fig:m1m0}
\end{figure}

As previously mentioned, the investigation of ISGMR in nuclei provides an important source of information on $K_\infty$. It is noteworthy that overall consideration for the calculations that reproduce the ISGMR in $^{208}$Pb and $^{90}$Zr by the non-relativistic and relativistic models, as well as the effects of the nuclear-matter symmetry energy, the $K_\infty$ should be located within the range of $240\pm20$ MeV \cite{Shlomo2006EPJA30.23, Garg2018PPNP101.55}. Combing the advantage of the TD-RHF model, the ISGMR mode of $^{208}$Pb is calculated using twelve relativistic effective Lagrangians. These include the RHF Lagrangians PKO$i$($i$=1,2,3) \cite{Long2006PLB640.150, Long2008EPL82.12001}, and the RMF ones  DD-ME1 \cite{Niksic2002PRC66.024306}, DD-ME2 \cite{Lalazissis2005PRC71.024312}, PKDD \cite{Long2004PRC69.034319}, TW99 \cite{Typel1999NPA656.331},  NL$i$ ($i$=1,2,3) \cite{Reinhard1986ZPA323.13, Lee1986PRL57.2916, Lalazissis1997PRC55.540}, NLSH \cite{Sharma1993PLB312.377} and PC-PK1 \cite{Zhao2010PRC82.054319}.

Figure \ref{Fig:m1m0} shows the constrained energy $\sqrt{m_1/m_{-1}}$ (MeV) extracted from the strength functions and the incompressibility $K_\infty$ given by selected models. The linear correlation between the ISGMR energies and $K_\infty$ is a commonly employed method to constrain the value of $K_{\infty}$ \cite{Blaizot1995NPA591.435}. As illustrated in Fig. \ref{Fig:m1m0}, there is a clear linear correlation between the constrained energy $\sqrt{m_1/m_{-1}}$ and $K_\infty$. The Pearson's coefficient reads as $0.974$, which is close to $1$. The early constraint of $K_\infty$ spanning from 250 to 270 MeV \cite{Ma2001NPA686.173, Vretenar2003PRC68.024310} is significantly reduced by the TD-RHF calculations in this work, which falls within 237$\sim$246 MeV and shows a good agreement with the range of 240$\pm$20 MeV \cite{Shlomo2006EPJA30.23, Garg2018PPNP101.55}. This reduction can be attributed to the fact that our calculation using the TD-RHF model can incorporate a more extensive range of popular effective Lagrangians, including the RMF and RHF ones, which ensures better systematics.

Furthermore, it is observed that the RHF Lagrangian PKO2, despite sharing a similar $K_\infty$ value with PKO1, gives a larger constrained energy. This significant difference may be primarily due to the deviation in nuclear-matter symmetry energy $J$. The $J$ values given by PKO2 and PKO1 read as 32.49 MeV and 34.37 MeV, respectively. The investigation of RMF+RPA and Skyrme-Hartree-Fock+RPA models indicates that the ISGMR energy tends to decrease with the enhancement of $J$ \cite{Vretenar2003PRC68.024310, Colo2004PRC70.024307}. This is consistent with the relatively large constrained energy given by PKO2, as compared to PKO1. Besides the $J$, the slope of the symmetry energy $L$ also influence the description of the energy of ISGMR mode \cite{Todd-Rutel2005PRL95.122501}. The investigations about $J$ and $L$ based on TD-RHF model will be carried out later.

\section{Summary}\label{sec:sum}

This work establishes the spherical time-dependent relativistic Hartree-Fock (TD-RHF) model by utilizing the spherical Dirac Woods-Saxon (DWS) base. The formalism of the time-dependent RHF equations based on the spherical DWS base is presented in detail. To illustrate the reliability of the time-evolution, the ISGMR mode of $^{208}$Pb is studied as an example. The high quality of the conservation of total binding energy and particle number, as well as the time-reversal invariance in the TD-RHF calculations, ensures the precision and reliability of our results and validate the description of the nuclear ISGMR mode.

A preliminary investigation into the ISGMR mode of $^{208}$Pb, which is commonly employed to constrain the incompressibility of nuclear matter $K_\infty$, has been conducted using the spherical TD-RHF model based on the spherical DWS base. The RHF Lagrangian PKO1 has been utilized as an example in this study. The results demonstrate that the constrained energy  of the ISGMR calculated by PKO1 is in close agreement with the experimental data. Moreover, the value of $K_\infty$ is constrained to a range of 237$\sim$246 MeV by the TD-RHF model, which employs twelve relativistic effective Lagrangians. This value is notably lower than the previous constraints based on the relativistic models, and is consistent with the range of $240\pm20$ MeV from the microscopic theory.

\begin{acknowledgments}
The authors thank the fruitful discussions with Dr. Z. X. Ren. This work is partly supported by the Strategic Priority Research Program of Chinese Academy of Sciences under Grant No. XDB34000000, the National Key Research and Development (R\&D) Program under Grant No. 2021YFA1601500, the National Natural Science Foundation of China under Grant No. 12275111, the Fundamental Research Funds for the Central Universities lzujbky-2023-stlt01, and the Supercomputing Center of Lanzhou University.
\end{acknowledgments}

%
%


\end{document}